# Optical image decomposition and noise filtering based on Laguerre-Gaussian modes

**JIANTAO MA,[1],[†] DAN WEI,[1],[†], HAOCHENG YANG,[1] YONG ZHANG,[1,2,*] MIN XIAO[1,2,3]**

[1]*National Laboratory of Solid State Microstructures, College of Engineering and Applied Sciences, and School of Physics, Nanjing University, Nanjing 210093, China*
[2]*Collaborative Innovation Center of Advanced Microstructures, Nanjing University, Nanjing 210093, China*
[3]*Department of Physics, University of Arkansas, Fayetteville, Arkansas 72701, USA*
*\*Corresponding author: zhangyong@nju.edu.cn;*
†*These authors contribute equally to this work.*

**We propose and experimentally demonstrate an efficient image decomposition in the Laguerre-Gaussian (LG) domain. By developing an advanced computing method, the sampling points are much fewer than those in the existing methods, which can significantly improve the calculation efficiency. The beam waist, azimuthal and radial truncation orders of the LG modes are optimized depending on the image information to be restored. In the experiment, we decompose an image by using about 3×10$^4$ LG modes and realize a high-fidelity reconstruction. Furthermore, we show image noise reduction through LG domain filtering. Our results open a door for LG-mode based image processing.**

Light beams carrying orbital angular momentum (OAM) have promoted many fascinating techniques such as digital spiral imaging[1-3], object's azimuthal identification[4-6], rotating object's sensing[7,8] and optical image rotation[9]. These applications are mainly driven by the mapping between azimuthal angle and OAM[10]. Laguerre-Gaussian (LG) mode, characterized by an azimuthal index *l* and a radial index *p*, is one popular OAM-carrying beam[11]. The recent emergence of LG mode sorter could further extend the applications to various fields [12-14].

Particularly, LG modes form an orthogonal basis, which can be utilized to decompose an image [15,16]. Generally, over ten thousands of LG modes are required to reconstruct a high-quality image. It is still a major challenge to analyze LG modes of such large range (especially their radial indices). And the LG mode distribution for a given image strongly depends on the selected beam waist, which is another barrier to find an optimal solution for image decomposition.

In this letter, we propose a systematic approach to decompose an image with about 3×10$^4$ of LG modes. An optimal beam waist is critical to minimize the LG mode number needed for image reconstruction. After the beam waist is chosen, the azimuthal and radial truncation orders are decided by the image complexity and detector pixels. To ensure the decomposition accuracy, especially for the high-order LG mode, we adopt least square method to calculate the radial mode distribution. Furthermore, we demonstrate LG domain filtering to reduce the azimuthal noise, which shows the potential application in optical image processing.

An arbitrary optical image can be expressed as a coherent superposition of LG modes,

$$U(r,\theta,z) = \sum_{l,p} A_{l,p} LG_{l,p}(r,\theta,z) \quad (1)$$

where $A_{l,p}$ is a complex variable that defines the amplitude and relative phase of LG mode, $r$ is the radius, $\theta$ is the azimuthal angle, and $z$ is the propagation distance. Note that all LG modes are determined by the same beam waist.

To simplify the analysis, we consider the image expansion at the position of the beam waist (z=0). The field distributions of LG modes are given by [17]

$$LG_{l,p}(r,\theta) = \sqrt{\frac{2p!}{\pi(p+|l|)!}} \frac{1}{w_0} \cdot \left[\frac{r\sqrt{2}}{w_0}\right]^{|l|} \exp\left[\frac{-r^2}{w_0^2}\right] L_p^{|l|}\left(\frac{2r^2}{w_0^2}\right) \exp(il\theta) \quad (2)$$

where $w_0$ is the beam waist, and $l$ and $p$ are the azimuthal and radial indices, respectively. $L_p^{|l|}$ is the generalized Laguerre polynomial. According to Eqs. (1) and (2), to expand or reconstruct an optical image $U$ in the LG domain, one should decide the coefficient $A_{l,p}$, the beam waist $w_0$ and the mode expansion range.

The LG mode range is related to *l*, *p*, and $w_0$. As shown in Eq. (2), the azimuthal distribution of the LG mode is related to the helical phase as well as the *l* index. The azimuthal expansion of an image can be written as

$$B_l(r) = \int_0^{2\pi} U(r,\theta) \exp(-il\theta) d\theta = \sum_p A_{l,p} LG_{l,p}(r) \quad (3)$$

where

$$LG_{l,p}(r) = \sqrt{\frac{2p!}{\pi(p+|l|)!}} \frac{1}{w_0} \left[\frac{r\sqrt{2}}{w_0}\right]^{|l|} \exp\left[\frac{-r^2}{w_0^2}\right] L_p^{|l|}\left(\frac{2r^2}{w_0^2}\right).$$

$B_l(r)$ shows the radial distribution in different OAM subspaces, which can be understood as the *l* spectrum at different radius.

Numerical computation of the integral in Eq. (3) is generally time-consuming. Here, we exploit the Fourier relationship between angular position and OAM[10] and calculate Eq. (3) by FFT at each radius after transforming the image $U$ to polar coordinate. Considering an optical image recorded by a 512×512 detector with 50 μm pixel size, the largest circumference contains 1609 pixels. Hence, the radial and azimuthal pixels of the target image are 256 and 1609, respectively (Fig. 1(a)).

To ensure high fidelity, we decide the truncation $l$ order by the following relation,

$$\sum_{l=-m}^{m} P_m(r) = 0.99 P(r) \quad (4)$$

$P(r)$ is the total power at radius $r$ and $P_m(r)$ is the power of the $m$-order $l$ component at radius $r$. Eq. (4) indicates that the azimuthal expansion is truncated when the $l$ spectrum from $-m$ order to $m$ order contains 99% power at the corresponding radius. The calculated truncation $l$ order for each radius is shown in Fig. 1(b). One can observe that the truncation $l$ order increases as the radius increases. The highest order is 105 at $r$=12.8 mm. It should be noted that, to analyze the $p$ components in $B_l(r)$, the range of the $l$ spectrum at each radius should be consistent. Thus, we extract the OAM components between ±105 orders to perform the azimuthal expansion. The actual azimuthal expansion accuracy of the whole image is more than 99% in the experiment.

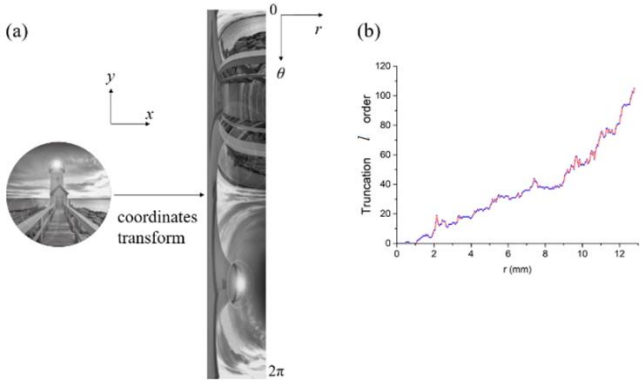

Fig. 1. (a) Image transformation from Cartesian coordinates to polar coordinates. (b) The azimuthal truncation $l$ order at each radius.

For an arbitrary $l$ order, $B_l(r)$ can be seen as a coherent combination of many $p$ components. Generally, an integral method can be used to calculate the weights of the $p$ components [15,16], in which a large sampling number is required to ensure high accuracy. Here, we propose a method to acquire the mode coefficient with a significantly reduced sampling number.

We regard Eq. (3) as a multivariate linear equation with different $r$:

$$\begin{cases} B_l(r_1) = A_{l,0} LG_{l,0}(r_1) + A_{l,1} LG_{l,1}(r_1) + \cdots + A_{l,p} LG_{l,p}(r_1) \\ B_l(r_2) = A_{l,0} LG_{l,0}(r_2) + A_{l,1} LG_{l,1}(r_2) + \cdots + A_{l,p} LG_{l,p}(r_2) \\ \vdots \\ B_l(r_m) = A_{l,0} LG_{l,0}(r_m) + A_{l,1} LG_{l,1}(r_m) + \cdots + A_{l,p} LG_{l,p}(r_m) \end{cases} \quad (5)$$

where $r_1, r_2 \ldots r_m$ are different radius. There are $p+1$ unknown variables $A_{l,0}$, $A_{l,1}$,..., $A_{l,p}$ to be ascertained. In our method, the minimal radial sampling number is $p+1$ to decide the radial mode coefficient; this is much smaller than that in integral method [15,16]. In our calculations, we sample $B_l(r)$ at different radius and solve the radial component by least squares fitting. To demonstrate the advantage of this fitting method, we analyze several LG mode fields ($p$=8 in Fig. 2(a), $p$=50 in Fig. 2(b), and $p$=9, 15, 28 superposed with 1:2:1 amplitude ratio in Fig. 2(c)). Note that LG modes with $l$=0 are used since we focus on the radial components here. Figures 2(d)–2(f) present the corresponding decomposition accuracy. The least squares fitting method presents a superior performance especially with a limited sampling number.

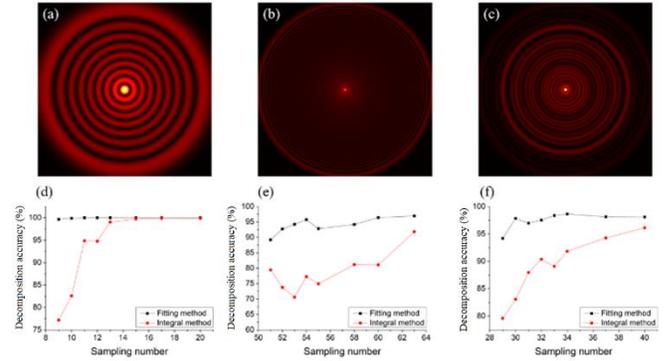

Fig. 2. Comparison of least squares fitting and the integral method. The target images are (a) a pure LG mode with $p$=8 and $w_0$=3000 μm, (b) a pure LG mode with $p$=50 and $w_0$=1260 μm, and (c) a superposed LG field with $p$=9, 15, 28 (1:2:1 amplitude ratio) and $w_0$=1700 μm. The decomposition accuracy in (d)-(f) corresponds to the fields in (a)-(c), respectively.

Generally, the used LG mode basis for expansion is strongly related to beam waist. To choose a suitable beam waist, one should have a clear knowledge of the radial feature of the LG beam. Figure 3(a) shows the radial distribution of the $LG_{l=5}^{p=50}$ mode with beam waist $w_0$. There are 50 nodes in the radial direction, and the distance between adjacent nodes gradually increases as the radius increases. The outer area, with an obviously larger node distance, offers limited spatial frequency for image decomposition. Thus, we remove the outermost eight nodes and define the remaining part as the effective area. Figure 3(b) shows $B_{l=5}(r)$, i.e., the $l$=5 subspace of the image to be decomposed. One can calculate the spatial domain width $r_{l=5}$ = 0.0126 m with 99% accuracy and the frequency domain width $f_{l=5}$ = 625 m$^{-1}$ with 95% accuracy.

To decompose the image, the effective area of the truncation LG mode must cover the radial distribution of the image and offer sufficiently high frequency. Therefore, the conditions of the truncation order for $p$ are:

$$N_1^{l,p} w_0 > r_l \quad (6a)$$

$$2\left(N_1^{l,p} - N_2^{l,p}\right) w_0 \leq 1/f_l \quad (6b)$$

where $N_1^{l,p}$ is the outermost node and $N_2^{l,p}$ is the second outermost node of the effective area of LG mode. $w_0$ should be determined so that Eqs. (6a) and (6b) are both satisfied in all OAM subspaces.

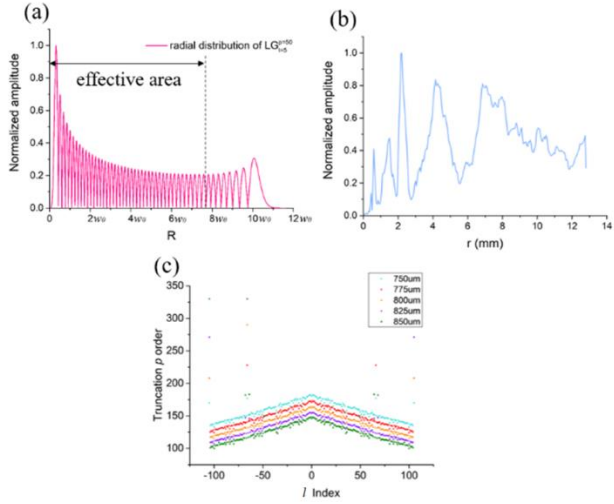

Fig. 3. (a) The radial distribution of LG mode ($l$=5, $p$=50) and its effective area. (b) The radial distribution of OAM subspace of $l$=5, (c) The truncation $p$ order with different beam waist.

Figure 3(c) shows the calculated truncation $p$ order (i.e., the smallest $p$ that satisfies Eq. (6a) and Eq. (6b)) with different beam waist. A larger beam waist requires a lower truncation $p$ order in general. However, the truncation $p$ order in some OAM subspaces would be extremely high for large beam waist. For example, when the beam waist is 800 μm, the truncation $p$ order in the $l$=±66 subspace is 290, which indicates a sampling number of at least 291 to solve all the radial components by using least square fitting method. However, the maximal sampling number is 256 due to the limitation of the detector in our experiment. Taking the above factors into consideration, we choose 775 μm as the optimal beam waist.

By applying least square fitting, one can obtain the amplitude and phase LG spectrum of the image as shown in Figs. 4(a) and 4(b), respectively. The total LG mode number is 31389 and the reconstructed image is shown in Fig. 4(c). The fidelity is defined as

$$F = \left| \frac{\iint I_r(x,y) \cdot [I_0(x,y)]^* dxdy}{\sqrt{\iint |I_r(x,y)|^2 dxdy \iint |I_0(x,y)|^2 dxdy}} \right|^2 \quad (7)$$

where $I_r$ and $I_0$ are the intensities of the reconstructed and original images, respectively. The fidelity is calculated to be 99.35%, which proves a good accuracy. The difference between the original and reconstructed images (Fig. 4(d)) is mainly caused by the energy loss from mode truncation.

Linear CCD rotary scanning is a popular imaging method. However, high-frequency dark noise of the detector could introduce an azimuthal periodic noise, which has a severe influence on the recorded image (Fig. 5(a)). It is convenient to filter such noise in LG mode domain because the major information for a natural image is carried by low-order LG modes. Figure 5(b) shows the $l$ spectrum of the scanned image. The periodic noise exists near ±180-order. By using the $l$ components ranging from -150 to 150, the recomposed image (Fig. 5(c)) clearly exhibits an improved quality.

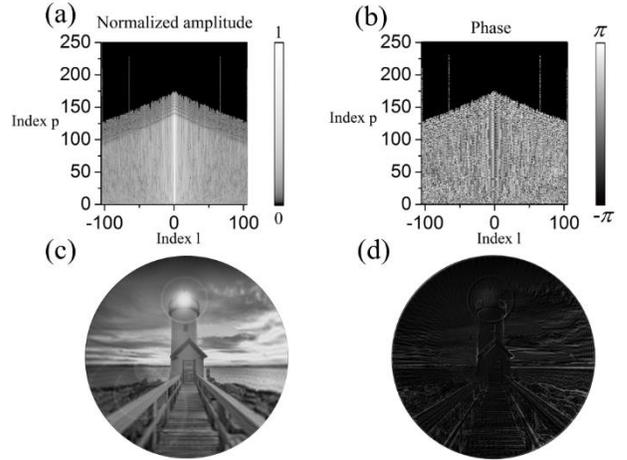

Fig. 4. (a) The amplitude of LG mode spectrum. (b) The phase of LG mode spectrum. (c) The reconstructed image by using the LG mode spectrum in (a) and (b). (d) The difference between the reconstructed and original images.

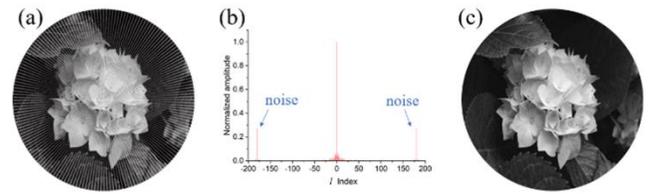

Fig. 5. (a) Raw image with azimuthal noise recorded by linear CCD. (b) The corresponding $l$ spectrum. (c) The reconstructed image after LG domain filtering.

In conclusion, we have introduced a systematic approach to decompose an optical image based on LG modes. The LG mode range for decomposing the image is related to the beam waist. We have developed an effective fitting method to analyze the radial mode component, which can significantly reduce the sampling number. In addition, we experimentally demonstrate the azimuthal noise filtering in LG domain. Our results pave a way for utilizing LG-mode basis in image processing.

**Funding.** National Key R&D Program of China (2017YFA0303703 and 2016YFA0302500), the National Natural Science Foundation of China (NSFC) (91950206 and 11874213), Natural Science Foundation of Jiangsu Province (BK20180322) and the Fundamental Research Funds for the Central Universities (1480605201)

**Acknowledgment** We acknowledge the authors of "Multiprecision Computing Toolbox", which is used for numerically evaluate the high-order LG mode.

**Disclosures.** The authors declare no conflicts of interest.